\documentclass[final,1p,times]{elsarticle} 
\usepackage{graphicx} 
\usepackage{amssymb} 
\usepackage{amsthm} 
\usepackage{lineno} 

\def\MeV {\mathop{\hbox{MeV}}}

\def\DU  {\mathop{{\cal D}\hbox{U}}}

\newcommand\detn[1]{\mbox{det}_{#1}}

\newcommand{\beq}{\begin{equation}}
\newcommand{\eeq}{\end{equation}}
\newcommand{\beqa}{\begin{eqnarray}}
\newcommand{\eeqa}{\end{eqnarray}}

\journal{Nuclear Physics A} 
\begin{document} 

\begin{frontmatter} 


\title{Study of QCD critical point using canonical
ensemble method}

\author{
\centerline{{\bf Anyi~Li}$^{\rm a}$ {\bf Andrei Alexandru}$^{\rm b}$
{\bf  Xiangfei Meng} $^{\rm c}$ and {\bf Keh-Fei Liu}$^{\rm a}$,
for the $\chi$QCD Collaboration}
}

\address{
\centerline{$^{\rm a}$ Department of Physics and Astronomy,
University of Kentucky, Lexington KY 40506, USA}

\centerline{$^{\rm b}$ Physics Department, The George Washington
University, Washington, DC 20052, USA}

\centerline{$^{\rm c}$ Department of Physics, Nankai University, Tianjin 300071, China}
}

\begin{abstract} 
The existence of the QCD critical point at non-zero baryon density is not
only of great interest for experimental physics but also a challenge for the
theory. We use lattice simulations based on the canonical ensemble
method to explore the finite baryon density region and look for the critical point.
We scan the phase diagram of QCD with three degenerate quark flavors using clover fermions
with $m_\pi \approx 700\mbox{MeV}$ on $6^3\times4$ lattices.
We measure the baryon chemical potential as we increase the density and we see the characteristic ``S-shape''
that signals the first order phase transition. We determine the phase boundary by Maxwell construction
and report our preliminary results for the location of critical point.

\end{abstract} 

\end{frontmatter} 



\section{Canonical ensembles}
The search for the QCD critical point has attracted
considerable theoretical and experimental attention recently.
To simulate matter at high density, a method based on the canonical
partition function was proposed~\cite{Liu:2003wy}. While expensive --
every update involves the evaluation of the fermionic determinant --
finite density simulations based on this method proved feasible~\cite{Alexandru:2005ix}.

We construct canonical partition function as a Fourier transform
of grand canonical partition function with respect to a $U(1)$ phase in the last time slice:
\beq
 Z_C(V, T, k) \equiv \int \DU e^{-S_g(U)} \detn{k} M^2(U)
\eeq
where $\detn{k}M^2(U)$ is the projected determinant on the $k$ quark sector.
With the aid of winding number expansion method~\cite{Meng:2008hj}, this can be computed efficiently. A program was
outlined to scan the QCD phase diagram to look for the critical point~\cite{Li:2006qa,Li:2007bj}. We refer readers to our
previous papers that include details on the simulations of the canonical partition function~\cite{Alexandru:2005ix}.

In this paper, we present results for $N_f=3$ based on simulations on
$6^3\times 4$ lattices with clover fermions. We fix temperature and scan
in the baryon number direction. By taking the difference
of free energy after adding one baryon, baryon
chemical potential can be measured as an observable in
canonical ensemble. We plot the
chemical potential as a function of baryon density. In a
finite volume, due to the non-zero contribution from
the surface tension, the first order phase transition will
be reflected as an ``S-shape'' structure in this plot~\cite{deForcrand:2006ec}. The
phase boundaries of the coexistence region can be determined
by ``Maxwell construction''.  We observe a clear signal for the first order phase transition.
With the phase boundary determined at simulation temperatures by ``Maxwell construction'',
we located the critical point at the intersection of the extrapolated phase boundary lines.

\section{Results}

\begin{figure}[t]
\centering
\includegraphics[height=1.5in]{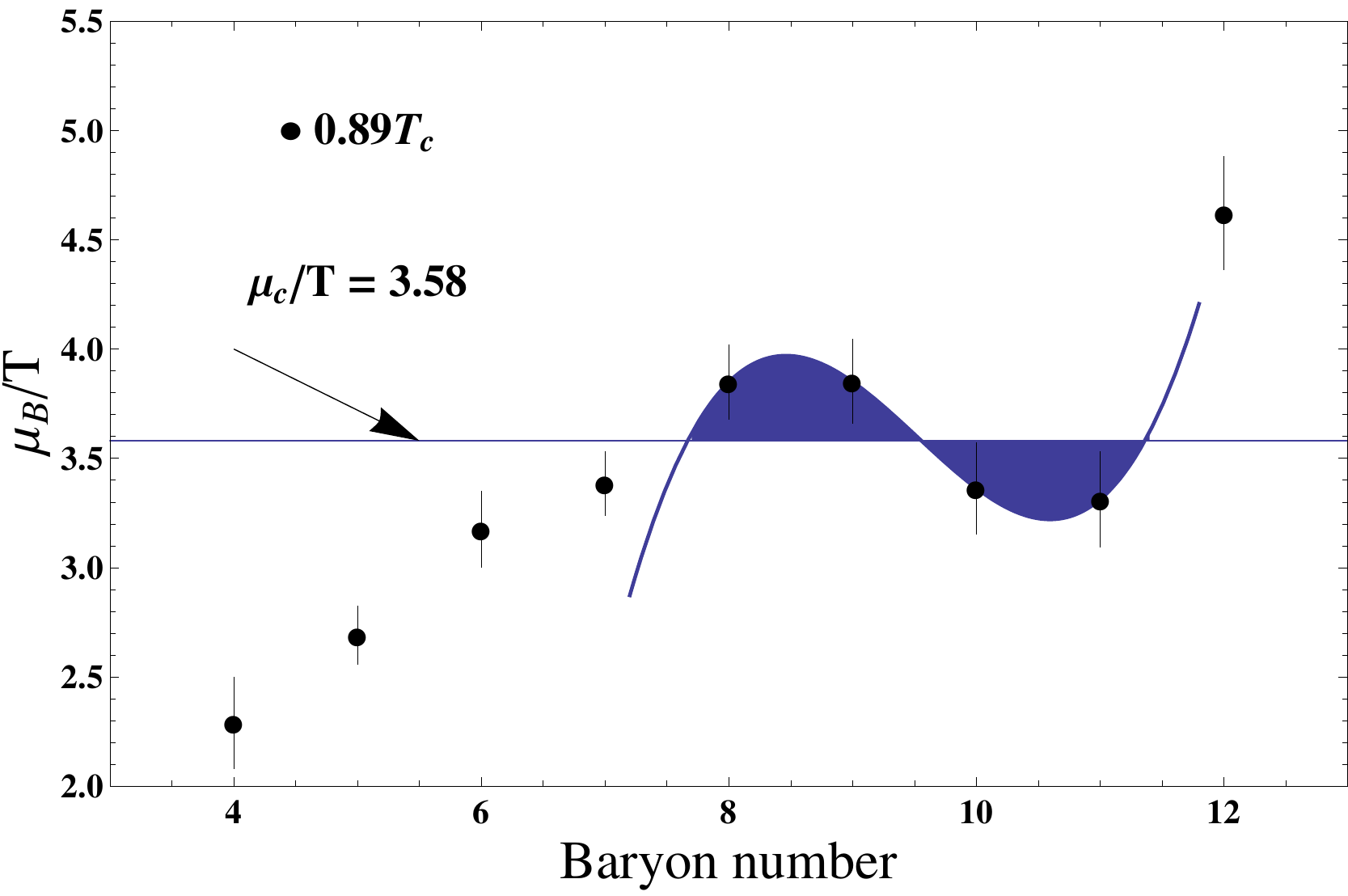}
\caption{``S-shape'' structure in the baryon chemical potential vs. baryon number plane as well as Maxwell construction}
\label{maxwell_construction}
\end{figure}

We illustrate the ``S-shape'' structure in Fig.~\ref{maxwell_construction} at a fixed temperature $T=0.83T_c$.
Once we determine the phase boundaries at a few temperatures, the critical point can be located
at the point where phase boundaries of coexistence phase cross together. Using the Maxwell construction
we also determine the value of the critical chemical potential and we plot the phase diagram in the $T-\mu_B$ plane.

\begin{figure}[h]
\centering
\includegraphics[scale=0.53]{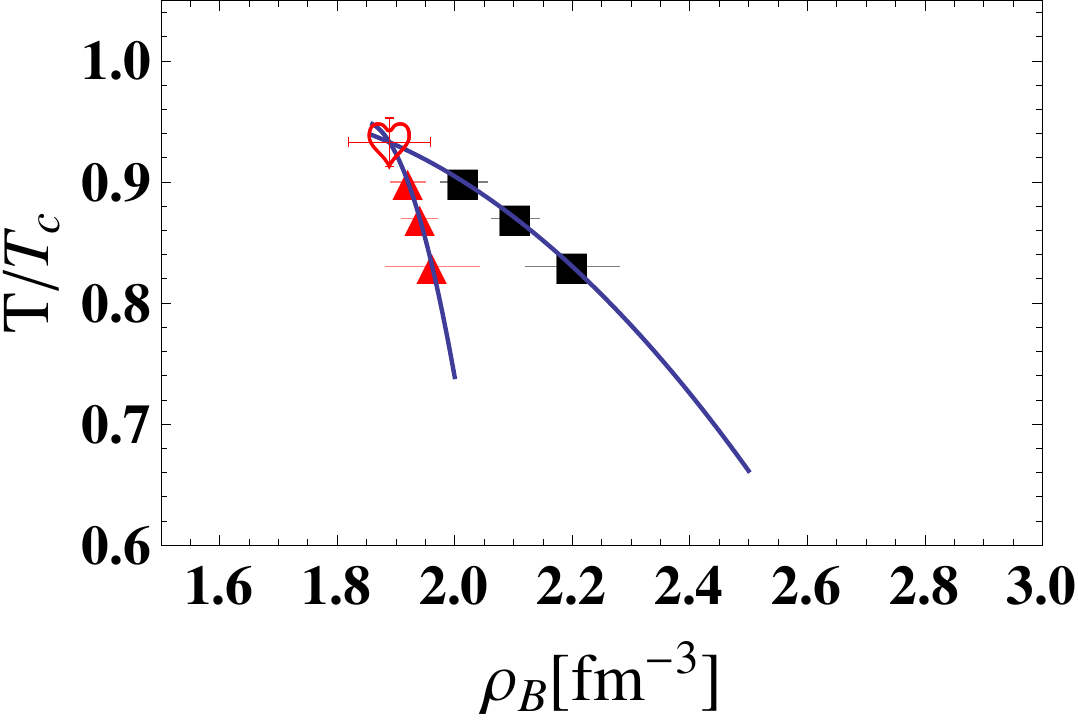}
\includegraphics[scale=0.53]{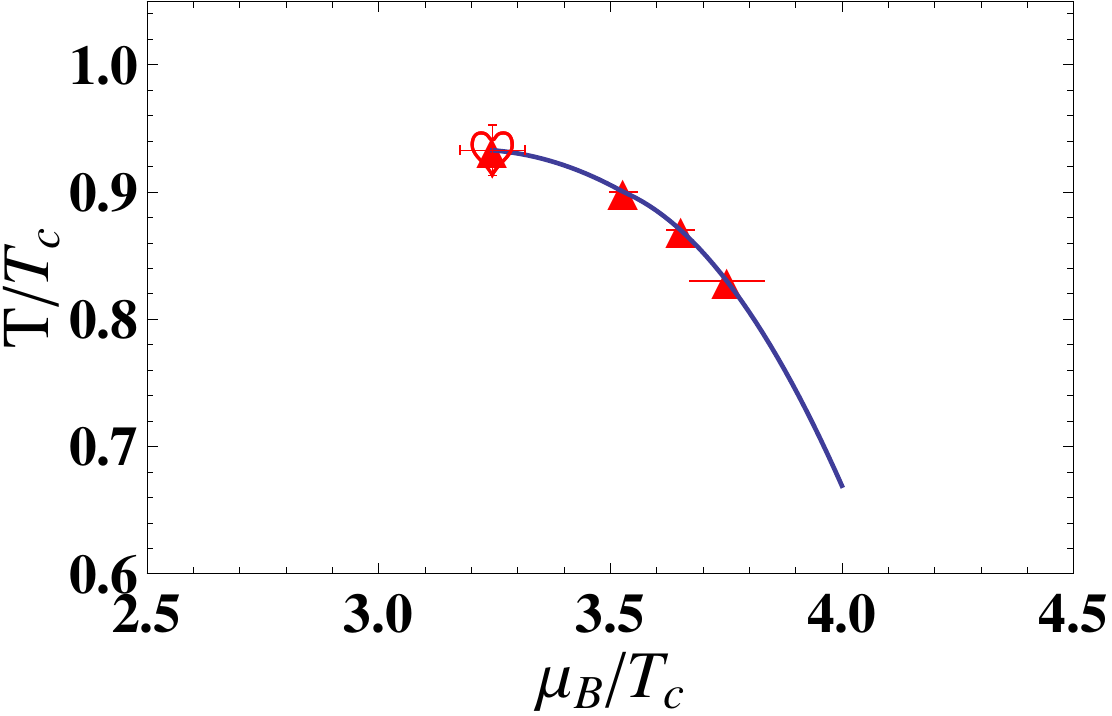}
\caption{Left panel: Boundaries of coexistence region.~~Right panel: Phase transition line in the $T-\mu_B$ plane. The critical point is located by an extrapolation}
\label{pdg}
\end{figure}

The critical point is found at $T_E/T_c = 0.93(2)$ and $\mu_B^E/T_c = 3.25(7)$ for $N_f=3$
and $m_\pi \approx 700\MeV$.


\section*{Acknowledgments} 
This work was supported in part by U.S. DOE grant DE-FG05-84ER40154. The calculation
was performed at Texas Advanced Computing Center (TACC) at The Univ. of Texas at Austin and The Univ. of Kentucky.

\end{document}